\documentclass[floatfix,superscriptaddress,aps,nofootinbib,prb,twocolumn,longbibliography]{revtex4-2}
\usepackage[utf8]{inputenc}
\usepackage{amsmath}
\usepackage{amsfonts,dsfont}
\usepackage{amssymb,amsthm}
\usepackage{graphicx}
\usepackage[margin=2cm]{geometry}
\usepackage[]{hyperref}

\hypersetup{
  colorlinks = true,
  urlcolor = Blue,
  linkcolor = red,
  citecolor = cyan
}

\usepackage{epsfig,color}
\usepackage{graphicx}
\usepackage{dcolumn}% Align table columns on decimal point
\usepackage{bm}% bold math
\usepackage{amsmath, amsfonts, amssymb,mathrsfs}
\usepackage{pstricks}
\usepackage{amsxtra}
\usepackage{amsthm}
\usepackage{natbib}
\usepackage{qcircuit}
\usepackage{array}

\AtBeginDocument{\tabcolsep=7pt}
\usepackage{booktabs}

\newcommand{\ket}[1]{\mbox{$|#1\rangle$}}
\newcommand{\bra}[1]{\mbox{$\langle#1|$}}

\def\be{\begin{equation}}      % with numbering
\def\ee{\end{equation}}
\def\beu{\begin{equation*}}   % without numbering
\def\eeu{\end{equation*}}

\DeclareMathOperator{\trace}{Tr}      % Trace of matrix
\theoremstyle{definition}

 \definecolor{new}{rgb}{.08,.05,.8}
\newcommand{\delete}[1]{{}} 

\begin{document}
\title{Compressed gate characterization for quantum devices with time-correlated noise}
\author{M. J. Gullans}
\affiliation{Joint Center for Quantum Information and Computer Science,
NIST/University of Maryland, College Park, Maryland 20742, USA}
\author{M. Caranti} 
\affiliation{Department of Physics, Princeton University, Princeton, New Jersey 08544, USA}
\author{A. R. Mills} 
\affiliation{Department of Physics, Princeton University, Princeton, New Jersey 08544, USA}
\author{J. R. Petta}
\affiliation{Department of Physics and Astronomy, University of California -- Los Angeles, Los Angeles, California 90095, USA}
\affiliation{Center for Quantum Science and Engineering, University of California -- Los Angeles, Los Angeles, California 90095, USA}

\begin{abstract}
As quantum devices make steady progress towards intermediate scale and fault-tolerant quantum computing, it is essential to develop rigorous and efficient measurement protocols that account for known sources of noise.   
Most existing quantum characterization protocols such as gate set tomography and randomized benchmarking assume the noise acting on the qubits is Markovian. However, this assumption is often not  valid, as for the case of $1/f$ charge noise or hyperfine nuclear spin noise.
%Many platforms suffer from non-Markovian noise sources that limit the  applicability of standard approaches for quantum process characterization.  
Here, we present a general framework for quantum process tomography (QPT) in the presence of time-correlated noise.   
We further introduce fidelity benchmarks that quantify the relative strength of different sources of Markovian and non-Markovian noise.
As an application of our method, we perform a comparative theoretical and experimental analysis of silicon spin qubits. %which leads to a number of results of independent interest for this platform.
We first develop a detailed noise model that accounts for the dominant sources of noise and validate the model against experimental data. 
Applying our framework for time-correlated QPT, we find that the number of independent parameters needed to characterize one and two-qubit gates can be compressed by 10x and 100x, respectively, when compared to the fully generic case.  
These compressions reduce the amount of tomographic measurements needed in experiment, while also significantly speeding up numerical simulations of noisy quantum circuit dynamics compared to time-dependent Hamiltonian simulation.  
Using this compressed noise model, we find good agreement between  our theoretically predicted process fidelities and  two qubit interleaved randomized benchmarking fidelities of $99.8\%$ measured in recent experiments on silicon spin qubits.  
More broadly, our formalism can be directly extended to develop efficient and scalable tuning protocols for high-fidelity control of large-arrays of quantum devices with non-Markovian noise.
\end{abstract}

\maketitle

\section{Introduction}
Accurately and efficiently characterizing noise in current quantum computing hardware is essential in realizing their long-term technological promise \cite{Martinis15}.  Many platforms suffer from non-Markovian, time-correlated noise processes such as drift and $1/f$ noise whose variance diverges with averaging time.  On the other hand, the standard assumption in quantum characterization protocols is that noise-processes have short range correlations in time with a finite variance \cite{Nielsen21}.  For example, the underlying theory of randomized benchmarking (RB) and gate set tomography (GST) explicitly neglect time-dependent fluctuations in the noise model.  
This assumption is referred to as the Markovian approximation because it arises from modeling the bath as a memory-less system in thermal equilibrium \cite{Breuer16}.  

A number of theoretical studies have made progress in extending quantum characterization methods to deal with different types of time-correlated noise, providing clear evidence of the need to account for these effects when comparing to experiment  \cite{Epstein14,Ball16,Fong17,Dehollain16,Mavadia18,Proctor20}.   However, these works have provided limited guidance on the experimental resources required to characterize non-Markovian noise.
At the same time, significant effort has been devoted to speeding up standard gate characterization protocols \cite{Kelly14,Barends19,Nielsen21,Brieger21}.  In the most general case, accounting for non-Markovian effects significantly increases the amount of characterization needed to describe quantum circuit dynamics \cite{White22}. As a result, compressed descriptions become especially important when dealing with correlated noise processes \cite{White21}.

Non-Markovian noise is particularly prominent in solid-state systems such as superconducting qubits and spin qubits.
In superconducting qubits, time-correlated noise arises primarily from electric field fluctuations in the environment \cite{Dial16}. The theoretical analysis of these effects in superconducting qubits have focused primarily on RB \cite{Epstein14,Ball16,Fong17}. 
In spin qubits, where quantum information is encoded in an electron or nuclear spin, the noise has a rich variety of non-Markovian sources including  lattice nuclear spins, other local magnetic field inhomogeneities, and spin-orbit coupling to electric field fluctuations \cite{Yoneda17,Burkard23}.  
Moreover, the first demonstrations of high-fidelity few-qubit control in spin qubits have just recently been obtained with two-qubit gate fidelities  $>99.8\% $ \cite{Xue22,Noiri22,Mills22}.  These results motivate a more detailed investigation of non-Markovian noise effects for quantum process tomography (QPT) of spin qubits.

In this article, we introduce a  method for QPT with time-correlated noise.  
Our basic approach to the problem is to model each quantum gate as a time-dependent quantum channel whose matrix logarithm, the so-called ``error generator,'' follows a (potentially non-stationary) Gaussian stochastic process \cite{Blume22}.  Our decomposition of the error channel allows us to naturally separate noise contributions into sources with different spectral characteristics.   The spectrally-resolved error generators then also lead to natural fidelity benchmarks for different sources of noise in the system.

As an application of our methods, we  perform a systematic analysis of silicon spin qubits, effectively providing a case study, while also leading to a number of new insights for this platform. Of primary importance, we find that our formalism allows for a significant compression of the noise model both in terms of the degree of temporal correlations and the number of independent parameters.  
We analyze the problem by first introducing a detailed noise model for silicon spin qubits.  We validate this model by comparing to experimental data on conventional one and two-qubit control experiments, finding excellent agreement between theory and experiment.  
With the microscopic noise model established, we then apply our framework for time-correlated QPT to this platform.  Crucially, we find that the dominant noise contributions for one and two-qubit gates are coherent errors with fluctuations that are effectively static on the time-scale of  quantum circuits that last for 1 ms or less.  

The resulting simplification of the noise model allows for a large speed up in both experimental gate characterization times and  numerical simulations of noisy quantum circuits.  Using our compressed model, we compare gate fidelities obtained from interleaved RB \cite{Magesan11} against exactly computed process fidelities.  We find that interleaved RB accurately measures the error rate of two-qubit gates in silicon spin qubits to within a factor of two, consistent with past studies for superconducting qubits  \cite{Epstein14}.  
These results are in good agreement with the recent experimental measurements of two-qubit gate fidelities $>99.8\% $ in silicon spin qubits \cite{Mills22}.

The article is organized as follows. Section~\ref{sec:tcqpt} introduces a general framework for time-correlated QPT.  We use this analysis to introduce fidelity benchmarks for different noise sources.  Section~\ref{sec:spinqubits} describes a noise model for silicon spin qubits and compares its predictions to experimental data taken from similar devices as Ref.~\cite{Mills22}.   Section~\ref{sec:QPTspin} applies our methods for time-correlated QPT to silicon spin qubits using numerically simulated noisy time evolution with the noise parameters extracted in Sec.~\ref{sec:spinqubits}.  We then use the tomographic results to develop a simplified error model for each gate, which is used to compare interleaved RB fidelities with directly computed process fidelities.  Section~\ref{sec:disc} provides some concluding remarks and the outlook for future work.

\section{Time Correlated Quantum Process Tomography}
\label{sec:tcqpt} 

We now introduce our general framework that allows us to incorporate time-correlated noise in the process tensors for quantum control.  We also define fidelity benchmarks for time-correlated noise that allow one to isolate different contributions to the gate infidelities from the measured data.

\subsection{Background and Setup}

In the Markovian approximation, each unitary gate operation is treated as a fixed quantum channel \cite{Nielsen21,Blume22}
\be
\mathcal{G}_i(\rho) = \mathcal{E}_i \circ \mathcal{U}_i (\rho),
\ee
where $\mathcal{U}_i (\cdot) = U_i \cdot U_i^\dag $ is the ideal gate operation and   $\mathcal{E}_i$ is a completely positive, trace preserving (CPTP) error channel that accounts for the noise in the gate.  Similarly, state preparation and measurement  (SPAM) errors are modeled through fixed quantum channels applied before and after the operation, respectively.  

Here, we adapt the standard theory of quantum characterization techniques in the presence of Markovian noise to account for broad spectrum noise fluctuations in the error channels for each gate.  To simplify the analysis and avoid overparameterization, we use a model that is based on a minimal extension of the Markovian case to treat time-correlated classical noise; however, it is worth noting that our framework could be further generalized to include a quantum environment following the recently introduced process tensor tomography framework \cite{White22}.  In our approach, we take each qubit to be coupled to spatially local noise fields with different spectral fluctuations.  For example, consider the Hamiltonian for a single-qubit
\be
H = H_0(t) + H_M(t) +  H_s(t) B(t) + H_f(t) V(t),
\ee
where $H_0(t)$ is the control Hamiltonian and $H_M(t)$ is a deterministic time-dependent error.  The third and fourth terms are proportional to fixed error Hamiltonians $H_s$ and $H_f$ that multiply random, zero-mean noise fields $B(t)$ and $V(t)$.   $B(t)$ is a random  field that models slowly varying magnetic field noise, while $V(t)$ has a power spectral density of $1/f$ noise that models electric field noise acting on the qubit.  We treat the field $B(t)$ as a ``quasistatic'', which means that it is taken as constant on the timescale of a full run of QPT, but changes on the timescale of minutes to hours.

In practice, developing and validating a full microscopic Hamiltonian model for the noise becomes difficult when considering the evolution of quantum circuits that involve multiple external, time-dependent drives.  Instead, we opt for a simplifying approximation that is equivalent to these more microscopic models to lowest order in the noise fields.  Specifically, we write the error channel for each $K$-qubit gate acting on qubits $\bm{n}=(n_1,\ldots,n_K)$ 
\beu
\log \mathcal{E}_i^{\bm{n}}(t) = L_{iM}^{\bm{n}}+\sum_k [L_{i s}^{\bm{n} k} B_{n_k}(t)  + L_{if}^{\bm{n} k} V_{n_k}(t)],
\eeu
where $t$ is the time at which the gate is applied, $B_{n_k}(t)$ and $V_{n_k}(t)$ are the local noise fields acting on qubit $n_k$, and $L_{i \mu}^{\bm{n} k}$ are the error generators for the gate that provide a simple generalization of a Hamiltonian to quantum channels \cite{Blume22}.  
We now show how to extract these error generators using a generalization of QPT.

\subsection{Tomographic Reconstruction}

To illustrate the general principles behind our approach, we first consider the case of a single qubit and neglect spatial correlations in the noise.  The results generalize in a straightforward manner to larger numbers of qubits, including the possibility of spatially correlated noise.
Given a tomographically complete set of initial states $\ket{\rho_i} \rangle$ and measurements $\ket{E_j }\rangle$, the error channel can be fully reconstructed from the measurement probabilities 
\be 
\label{eqn:pjki}
\begin{split}
p_{jk}^i(t) = \langle \bra{E_j} \mathcal{E}_i^n(t) \circ \mathcal{U}_i \ket{\rho_k} \rangle.
\end{split}
\ee
In experiment, we only obtain samples from these distributions.  As a result, it is impossible to obtain an instantaneous estimate of $\mathcal{E}_i^n(t)$ in a single shot with only a single copy of the system.

To model the measurement process, we let $x_{jk}^i(t)$ be a random variable that is equal to $1$ if the measurement outcome at time $t_i$ is $E_j$ and 0 otherwise.  A standard estimator for $p_{jk}^i$ is obtained from many sequential single-shot measurements
\be
\begin{split}
\hat{p}_{jk}^i(t,T,\delta t) &= \frac{\delta t}{T} \sum_n x_{jk}^i(t_n) \\
&\approx \frac{1}{T} \int_{t}^{t+T}dt'  \langle \bra{E_j} \mathcal{E}_i^n(t') \circ \mathcal{U}_i \ket{\rho_k}\rangle,
\end{split}
\ee
where $\delta t$ is the interval between measurements and $T$ is the total measurement time.  This averaging process effectively acts as a filter on the time-dependent fluctuations of $V_\ell (t)$ that cuts off frequencies $\omega \gg 1/T$.  As a result, we can write an effective model for $\hat{p}_{jk}^i(t,T,\delta t)$ with new random  fields $\hat{V} (t,T)$
\begin{align} 
\hat{p}_{jk}^i(t,T) &= \langle \bra{E_j} \hat{\mathcal{E}}_i^n(t,T) \circ \mathcal{U}_i \ket{\rho_k} \rangle, \\ \label{eqn:errgen}
\log \hat{\mathcal{E}}_i^n(t,T) &= L_{iM}^n+L_{i s}^n B_n + L_{if}^n \hat{V}_i(t,T),
\end{align}
where the Fourier transform of $\hat{V}_i(t,T)$ satisfies $\hat{V}_\ell (\omega) = o(1/\omega^2)$ for $\omega > 1/T$ and $\hat{V}_\ell(\omega) \approx V_\ell (\omega)$ for $\omega \ll 1/T$.

To extract the error generators, one first performs full tomography of $\hat{\mathcal{E}}_i^n$ on a timescale $T$ and then takes its matrix logarithm.    
Denoting averages over each noise source by $\mathbb{E}_a$, we have  $\mathbb{E}_s B_n = \mathbb{E}_f V_n(t,T) = 0$.  This property allows one to extract $L_{iM}^n$ by averaging the error generators over a long time interval.  Although, these noise terms have a contribution that averages to zero, they still contribute significantly to the error channel through their fluctuations.  By examining the power spectral density of the error generators, we can directly measure these fluctuations. More precisely,  the magnitude of the matrix elements of $L_{is/f}^n$ can be determined by examining the sample-averaged power-spectral density of individual matrix elements of $\log \hat{\mathcal{E}}_{i}^n(t) - L_{iM}^n$.  To determine the sign of the matrix elements we can use a microscopic model to estimate the sign of one of the error generator terms, which can then be used to fix the remaining signs.

\subsection{Fidelity Benchmarks for Time-Correlated Noise }

Many quantum devices suffer from a variety of fluctuating noise sources.  Their relative importance is usually quantified through Ramsey and spin-echo type measurements of the single qubits.  However, these metrics may not be reflective of the relative importance of the different noise contributions in the specific gate implementations.  Such detailed information on the individual gate performance is crucial for advanced error mitigation strategies \cite{McArdle20,Cai22} and noise-tailored fault-tolerant protocols \cite{Tuckett19}.      

One application of our formalism from the previous section is the ability to extract fidelity benchmarks for the different noise sources in Eq.~\eqref{eqn:errgen}.  We propose relative fidelity benchmarks as  generalizations of the average gate fidelity
\begin{align}
 F_{iM}^n &= \int d \psi \langle \bra{\psi} e^{L_{iM}^n} \ket{\psi} \rangle ,\\
F_{is}^n &=  \mathbb{E}_{s} \int d \psi \langle \bra{\psi} e^{L_{is}^nB_n}  \ket{\psi} \rangle ,\\
F_{if}^n &=  \mathbb{E}_{f} \int d \psi \langle \bra{\psi} e^{L_{if} \hat{V}_n(t) }\ket{\psi} \rangle,
\end{align}
and similarly for the different combinations of noise, e.g., $F_{iMs}^n =\mathbb{E}_s \int d \psi \langle \bra{\psi} e^{L_{iM}^n+L_{is}^n B_n} \ket{\psi} \rangle$.  The average gate fidelity is given by $F_{iMsf}^n$ in this notation.  These metrics serve as benchmarks for the relative importance of different noise sources for each gate, providing detailed information on  device improvements needed to enhance the performance of quantum circuits.

\subsection{SPAM Errors}

In the analysis above, we neglected the effects of SPAM errors on the error generators.  However,  our framework fits readily into the setting used in GST and RB that accounts for SPAM errors in a self-consistent manner.  To generalize our approach to the setting with SPAM errors, one would simply need to allow time-correlated noise in the state-preparation and measurement error generators.  The rest of the analysis presented here would then follow in a straightforward manner.

\section{Silicon Spin Qubit Noise Model}
\label{sec:spinqubits}

In this section, we develop a detailed noise model for correlated noise processes in silicon spin qubits and compare it to experimental data from devices similar to those used in Ref.~\cite{Mills22}.  The model explicitly includes time-correlated $1/f$-noise and quasi-static noise in the Hamiltonian parameters arising from electrical noise and nuclear spin fluctuations.  This model is used to benchmark our analysis of time-correlated QPT in the later sections.  A main result of this paper is to develop a compressed noise model (see Sec.~\ref{sec:trunc}) for noisy quantum circuit dynamics that quantitatively matches the behavior of the microscopic Hamiltonian model introduced in this section.

\begin{figure*}[t!]
\begin{center}
\includegraphics[width=.99\textwidth]{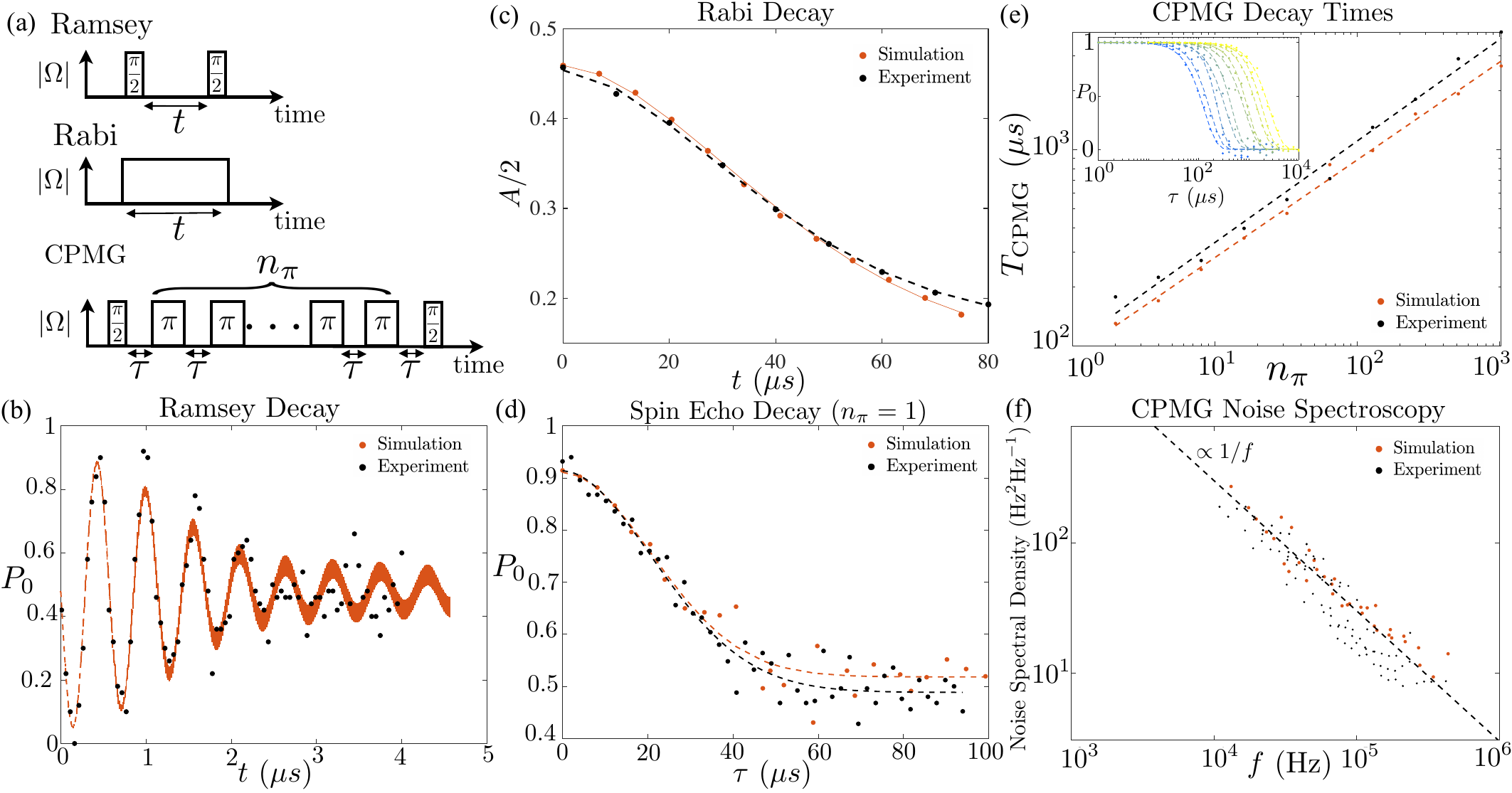}
\caption{Benchmarking the noise model: (a) Pulse sequences used in experiments.  The initial qubit state is spin-down. (b) Ramsey decay comparing simulation and experiment. $P_0$ is the spin-down return probability.  In (b)-(f), we took parameters $T_2^*=1.9~\mu$s, $T_2=40~\mu$s,~$f_c=10~$MHz, $f_\ell = 100$~Hz, and $\sqrt{A_0}=0.5~\mu$eV \cite{Mi18,Yoneda17}.  
(c) Decay of Rabi amplitude for a pulsed drive of a given time $t$. 
 In the simulation, we took $\Omega^0/2\pi=5$~MHz. $A$ is the amplitude of the Rabi oscillations in the spin-down probability at the $\pi$ times.  
(d) Spin echo decay comparing simulation and experiment.  The dashed lines are fits to a Gaussian. (e) CPMG decay times as a function of  sequence length $n_{\pi}$. Inset: Individual decay curves for the CPMG sequences. (f) Extracted spectral density from the CPMG data using the methods described in Ref.~\cite{Yoneda17}.  Both simulation and experiment show clear signatures of $1/f$ noise. 
 }
\label{fig:noise_bench}
\end{center}
\end{figure*}

\subsection{Model}

 The low-energy Hamiltonian for an array of silicon quantum dots with one electron per site takes the form \cite{Russ18}
\be
H= \sum_i g \mu_B \bm{B}_i^{\rm tot}\cdot \bm{s}_i +  \sum_{i,j} J_{ij} (\bm{s}_i \cdot \bm{s}_j - 1/4),
\ee
where $B_i^{\rm tot} = B_{\rm ext} \hat{z} + \bm{B}_i^M$ is the local magnetic field of spin $i$, including the contributions from a global external field and local fields, $s_i^\mu$ are spin-1/2 operators on site $i$, and $J_{ij}$ is the exchange interaction between dots $i$ and $j$, which are assumed to be in a quasi-1D spatial arrangement.  Single-site quantum control is achieved in this system using electric-dipole spin resonance (EDSR), while two-qubit gates are implemented through time-dependent control of the exchange interaction.   
\begin{align}
\label{eqn:Heff}
H &= \sum_i \hbar \Delta_i s_i^z + \sum_i \hbar \Omega_{i} (e^{-i \theta_i} s_i^+ + h.c. )\\
&+\sum_{i,j} J_{ij} (s_i^z s_j^z -1/4) + \frac{J_{ij}}{2} [e^{i(\omega_i - \omega_j)t} s_i^+ s_j^- + h.c.], \nonumber
\end{align}
where we have moved each spin into a local reference frame rotating with the frequency of the EDSR drives $\omega_i$ and neglected far-off resonant terms. The parameter $\Delta_i = g \mu_B B_i^{\rm tot}/\hbar - \omega_i$ is the detuning between the local field and the EDSR drive,  $\Omega_{i}$ is a real EDSR Rabi frequency, and $\theta$ is the axis of rotation.  

For sufficient  control over $\Omega_{i}(t)$, $\theta(t)$, and $J_{i,j}(t)$, this Hamiltonian is capable of universal quantum computation.  However, we also need to account for time-dependent fluctuations due to charge noise and nuclear spins.  We model these noise processes using the parameterizations \cite{Gullans19}
\begin{align}
\Delta_i(t) &= \Delta_i^0 +  \Delta^n_{i} v_{i}(t) +g \mu_B B_n/\hbar,  \\
\Omega_{i}(t) & = \Omega_{i}^0[1  +  \delta \Omega^n_{i} v_{i}(t)] ,\\
J_{ij}(t) & = J_{ij}^0 \{1 +  \delta J_{ij}^n  [v_i(t) + v_j(t)] \},
\end{align}
where $v_i(t)$ is a classical noise field on dot $i$, $\Delta_i^0$ is the target detuning of qubit $i$ from the EDSR frequency $\omega_{i}$, $B_n$ is a random Hyperfine magnetic field from the nuclear spins that we take to be static over the course of the full QPT experiment (but fluctuating between QPT runs on the timescale of minutes), $\Omega_i^0$ is the target EDSR Rabi frequency,  and $J_{ij}^0$ is the target exchange.    $\Delta_i^n$ is a noise sensitivity parameter that describes the change in the qubit frequency, $\delta \Omega_i^n$ measures the fractional change in the EDSR Rabi frequency,
and  $\delta J_{ij}^n$ measures the fractional change in the exchange interaction between qubits $i$ and $j$ in response to the noise, under the simplifying assumption that the exchange couples with equal magnitude to the noise field for each dot.   Finally, we remark that in experiments the term $\delta \Omega^n$ is expected to be quite small relative to the other effects.  Typically, the fluctuations in the Rabi frequency are instead driven by temperature drifts that occur on a slower timescale relative to the effects considered here.

We take the noise field $v_i(t)$ to have a $1/f$  power spectral density 
\begin{equation}
    S(f) = A_0/f,~f_\ell < f <f_c,
\end{equation}
where $A_0$ is the amplitude and $f_c$ and $f_\ell$ are high and low-frequency cutoffs.
For $f<f_\ell$ we take a white noise spectrum $S(f)=A_0/f_{\ell}$ and, for $f>f_c$, we have a cutoff of the form $S(f)=A_0 f_c/f^2$.
Neglecting spatial correlations in the noise, the noise sensitivity parameters can be related to the coherence times $T_{2i}^*$ of the qubits and the envelope decay rates of Rabi $\gamma_{ri}$ and exchange $\gamma_{e ij}$ oscillations \cite{Gullans19}
\begin{align}
\Delta_i^n & = \sqrt{\frac{1}{A_0 \log \big(\frac{f_c}{f_\ell} \big)}}\sqrt{ \frac{1}{T_{2i}^{*2}}- \frac{(g\mu_B  B_n)^2}{2\hbar^2}}, \\
\delta \Omega_i^n &=  \sqrt{\frac{1}{A_0 \log \big(\frac{f_c}{f_\ell}\big)  }}  \frac{\gamma_{ri}}{\Omega_i^0}, \\
\delta J_{ij}^n & = \sqrt{\frac{2}{ A_0 \log \big(\frac{f_c}{f_\ell}\big) }} \frac{\gamma_{e ij}}{J_{ij}^0}. 
\end{align}
  We can estimate $\Delta B_n$ from the value of $T_2$ assuming a large value of $\omega_c$ \cite{Ithier05}
\be
(g\mu_B  B_n/\hbar)^2 = \frac{2}{T_2^{*2}} - \frac{\log ( f_c/f_\ell)}{T_2^2 \log 4 }.
\ee

\subsection{Benchmarking the noise model with spin qubit experiments}
\label{sec:exp}

The noise model described in the previous section is much simpler than the most general possible noise model for two-qubits with time-correlated error generators.  Therefore, it is important to first test its predictions against more standard noise characterization protocols from the fields of nuclear magnetic resonance and electron spin resonance that also capture the presence of time-correlated noise processes \cite{Abragam1961,Slichter1996}.  The pulse sequences we employed to benchmark the noise model are shown in Fig.~\ref{fig:noise_bench}(a).

In Figs.~\ref{fig:noise_bench}(b)--(f), we compare the predictions of our noise model with four different single-qubit experiments: (i) Ramsey decay to measure $T_2^*$, (ii) Rabi oscillations to extract the quality factor of the Rabi oscillations, (iii)  spin-echo decay to measure $T_2$, and (iv) Carr-Purcell-Meiboom-Gill (CPMG) spectroscopy to extract the noise power spectral density of the qubit \cite{Bylander11,Medford12,Yoneda17,Connors22}.  
To perform this analysis we first compute the measured experimental quantities within our noise model.  We then did a nonlinear least-squares fit to extract estimates of the parameters of the noise model to model the experiment.  In all cases, we see good agreement between the simulations and experiment.    Most notably, we see clear evidence of the $1/f$ nature of the qubit noise in the CPMG spectroscopy \cite{Eng15,Yoneda17,Connors22}.

These simulations explicitly neglect incoherent processes such as $T_1$ decay and stochastic Pauli noise processes.  The neglect of $T_1$ decay is justified by the extremely long $T_1$ times of silicon spin qubits that routinely approach 100~ms \cite{Hayes09,Xiao10,Simons11,Yang13,Petit18}, including in the presence of a micromagnet \cite{Borjans19}.  This timescale should be compared to the time of a single gate, which is on the order of 50--200~ns.  As a result, $T_1$ decay will contribute to the error generators at the level of $10^{-6}-10^{-7}$.  The neglect of stochastic Pauli noise is based on a microscopic model for how dissipation arises in these systems.  It is worth noting that our model does give rise to stochastic Pauli noise from the high-frequency contributions to the $1/f$ charge noise.  However, we find in the following section that these effects provide a negligible contribution to the error rates for the gates in experimentally relevant parameter regimes.  Given the excellent agreement between this noise model and the experimental data it is a reasonable assumption to use our approximations as a baseline model.  A more careful experimental investigation of the stochastic error generators for one and two-qubit gates will be needed to more definitively rule out other sources of noise.

\section{Time Correlated QPT of Spin Qubits}
\label{sec:QPTspin}

In this section, we use the framework developed in Sec.~\ref{sec:tcqpt} to study the behavior of the error generators for one and two-qubit gates simulated using the noise model described in Sec.~\ref{sec:spinqubits}.  Using the information obtained from the error generators, we then develop a compressed model for gate noise in silicon spin qubits.  Using fast numerical simulations enabled by the compressed model, we then study performance of interleaved RB circuits with experimentally relevant parameters.

\subsection{Single-Qubit Gates}

To illustrate the basic principles behind our methodology for dealing with time-correlated noise, we first provide a detailed analysis of the error generator statistics  for single qubit gates.  

We consider a universal gate set consisting of $\pi/2$ rotations about the $X$ and $Y$ axes, plus $Z$ rotations with an arbitrary phase.  Since the $Z$-rotations are implemented in software when applying the $X$ and $Y$ rotations, they can be treated as having perfect fidelity \cite{McKay17}.   In the case of the $X$ and $Y$ rotations, the symmetries of our noise model imply that it is sufficient to just treat one rotation axis.  Therefore, we analyze the case of a single $\pi/2$ rotation about the $X$ axis in the Bloch sphere.

\begin{figure}[tb!]
\begin{center}
\includegraphics[width=.48\textwidth]{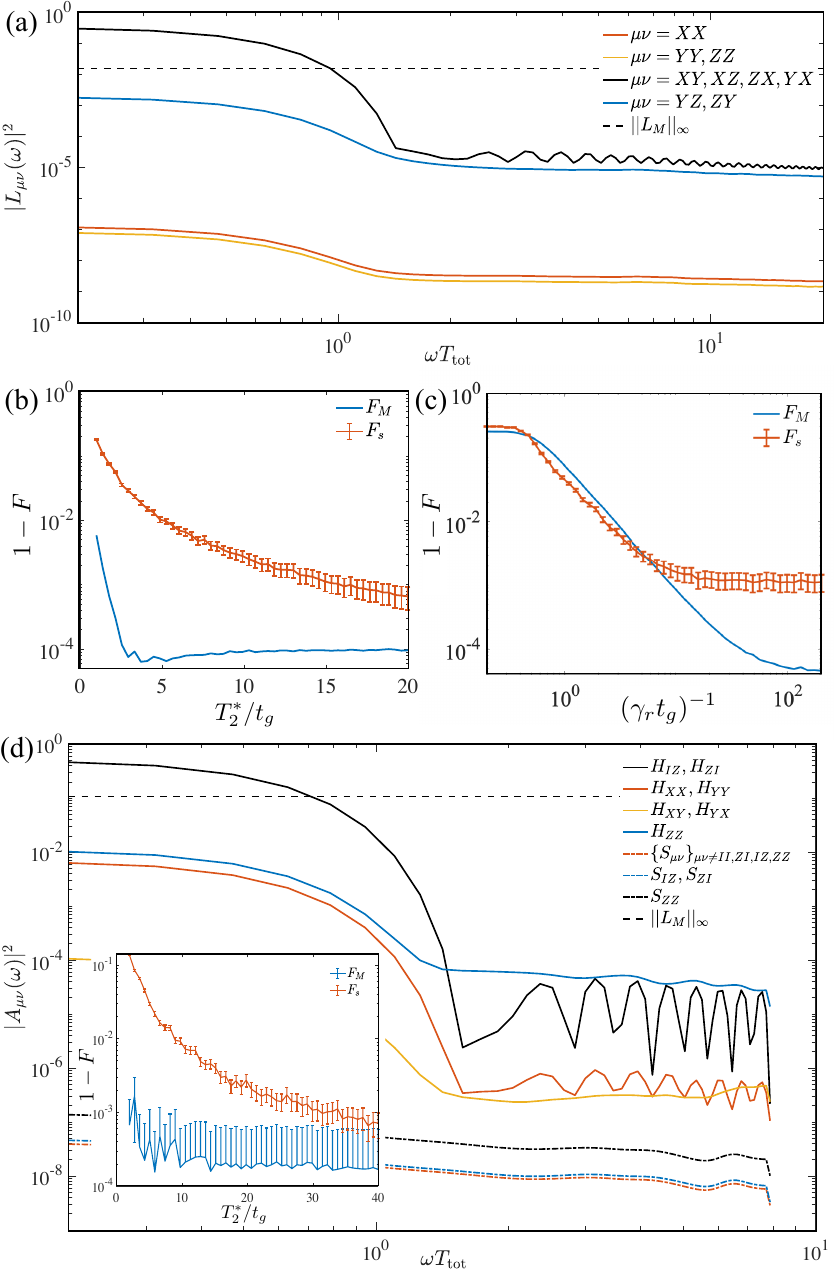}
\caption{(a) Power-spectral density of the matrix elements of the error generator for $\pi/2$ X-gate.  
We took $t_g = 100$~ns, $f_\ell= 100$~Hz, $f_c =10$~MHz, $\gamma_r=8$~kHz, $T_2^*=3~\mu$s, $T_2 = 30~\mu$s, and $T_{\rm tot}=1.6$~ms.  (b,c) Infidelity of Markovian $F_M$ and quasistatic $F_s$ noise sources as a function of $T_2^*$ and $1/\gamma_r$, respectively.  We took $\gamma_r=40$~kHz in (b) and $T_2^* = 1.5~\mu$s in (c).  (d) Power-spectral density of the matrix elements of the error generator for a two-qubit CZ gate.   We took $t_g = 50$~ns, $J_0=10$~MHz,  $\gamma_e=45$~kHz, $T_2^*=0.5~\mu$s for both qubits, $T_2 = 30/105~\mu$s for the two qubits, and $T_{\rm tot}=0.05$~ms.  Inset: Infidelity of Markovian $F_M$ and quasistatic $F_s$ noise sources as a function of $T_2^*$ with other parameters fixed.}
\label{fig:errgen}
\end{center}
\end{figure}

We analyze the gate characteristics using numerical simulations of the gate implemented via square pulses in the presence of quasistatic and $1/f$-noise with noise parameters taken from Sec.~\ref{sec:exp}.  We generate the $1/f$ noise by generating iid complex Gaussian noise in frequency space with the appropriate variance in each frequency bin (as set by the power-spectral density) \cite{Timmer95}.  We then obtain $V(t)$ by Fourier transforming to real time.

From these simulations, we then study the gate process matrix at different starting times for the gate over the course of one time-dependent realization of the correlated noise.  To account for the average in Eq.~\ref{eqn:pjki} needed to estimate the gate parameters, we take a mixture of the noisy realizations of each gate over several hundred consecutive gate times that is needed to build up sufficient measurement statistics.  We then compute the error generators over time for up to several thousand total gate times and multiple time-dependent realizations of the correlated noise. In this way we can numerically sample the full spectral density of the error generators for the gate in a way that approximates a realistic experiment.  The results for the power spectral density of the individual matrix elements of the Pauli transfer matrix for the error generators are shown in Fig.~\ref{fig:errgen}(a).  Here, $t_g$ is the time of one gate and $T_{\rm tot}$ is the total length of time of each noise realization (i.e., we simulate $T_{\rm tot}/t_g$ consecutive implementations of the gate).   

Interestingly, we see from Fig.~\ref{fig:errgen}(a) that the higher frequency fluctuations of the error generators are strongly suppressed compared to the low-frequency dynamics.  These high-frequency contributions to the error generators are sufficiently weak that they can be safely neglected in analyzing the gate fidelities at the level of  error rates above $10^{-4}$.   Moreover, the error generators are seen to be dominated by low-frequency fluctuations of the off-diagonal terms corresponding to coherent errors.  The diagonal stochastic errors are strongly suppressed at all frequencies.  

To capture the role of the Markovian noise relative to the low-frequency fluctuations, we use a model for the error generator of the gate 
\begin{equation}
    L_{\rm eff} = L_M + L_s R,
\end{equation}
where $L_M = \mathbb{E}[ \log \mathcal{E}]$ for the error channel $\mathcal{E}$ for the gate,  $|L_{s\mu \nu}|^2 = \mathbb{E}[(L_{\mu \nu} - L_{M\mu \nu})^2] $, and $R$ is Gaussian random field with zero mean and unit variance.  This model combines the effects of $L_f$ with the static terms $L_s$.  This simplification is allowed because we find that the high-frequency components of the error generators have a negligible contribution to the overall error channel.
The relative sign of the matrix elements for $L_s$ are set by the signs of one realization.  With this approach to the parameterization, we  then compute the fidelity benchmarks $F_M$ and $F_s$ for this gate, see Fig.~\ref{fig:errgen}(b-c).  In the regime where the Rabi frequency suffers from weak fluctations $(\gamma_r t_g)^{-1} \gtrsim 100$, we can see from Fig.~\ref{fig:errgen}(b) that the Markovian terms are negligible compared to the quasistatic effects.  However, as the Rabi decay rate increases $(\gamma_r t_g)^{-1} \lesssim 10$, Fig.~\ref{fig:errgen}(c) shows that the Markovian contributions to the noise become comparable to the quasistatic contributions.  

\subsection{Two-Qubit Gates}

We now apply our formalism to the case of two-qubit exchange gates.  We focus on the two-qubit CZ-gate whose implementation is 
\be
U_{\rm CZ} = e^{i \pi (s_1^z+s_2^z)/2} e^{-i J_{12} s_1^z s_2^z t_{\rm ex}}
\ee
where $t_{\rm ex} = \pi/J_{12}$.  In the presence of a large magnetic field gradient, this gate can be implemented by adiabatically turning the exchange on and off, followed by a $\pi/2$-rotation about the $Z$ axis for both qubits \cite{Meunier11,Russ18,Watson18}.

 In the case of two-qubits, the number of individual matrix elements of the error generator is much larger (256 compared with 16 for a single qubit).  Therefore, to present a more compressed representation of the noise sources we use the formalism of Ref.~\cite{Blume22} to break up the error generator into contributions from different noise types: (i) Hamiltonian noise, (ii) stochastic noise, and (iii) active noise.  Hamiltonian noise gives a contribution to the error generator that we denote by $\mathcal{H}_P$ for a Pauli operator $P$.  This term acts on density matrices as
\begin{equation}
    \mathcal{H}_P[\rho]=-i  [P,\rho].
\end{equation}
The stochastic error generator term  $\mathcal{S}_P$ for a Pauli operator $P$ acts diagonally on the density matrix as \begin{equation}
    \mathcal{S}_P[\rho] =  P  \rho P^\dag - \rho.
\end{equation}  
  There are also two other types of error generators called symmetric and anti-symmetric generators that, for two Pauli operators $P$ and $Q$, take the form
\begin{align}
    \mathcal{C}_{P,Q}[\rho] &= P\rho Q + Q\rho P - \frac{1}{2}\{\{P,Q\},\rho \},\\
    \mathcal{A}_{P,Q}[\rho] & = i\Big(P\rho Q - Q \rho P -\frac{1}{2} [ \{P,Q\},\rho ] \Big).
\end{align}
For each of these types of terms, there is a corresponding error rate associated with its contribution to the error generator.  To calculate the error rate we introduce a dual basis 
\begin{align}
\mathcal{H}_P'&=\mathcal{H}_P/d^2, \\
\mathcal{S}_P'[\rho] &= P \rho P^\dag/d^2, \\
\mathcal{C}_{P,Q}'[\rho] &= (P \rho Q + Q \rho P)/2d^2,\\
\mathcal{A}_{P,Q}'[\rho] &= i(P \rho Q - Q \rho P)/2d^2,
\end{align}
where $d$=4 is the Hilbert space dimension for two qubits.
The dual basis was introduced because it satisfies the identity
\begin{equation}
    \trace[\mathcal{B}'^\dag \mathcal{D}] = \delta_{\mathcal{B}' \mathcal{D} },
\end{equation}
where $\mathcal{B}'$ and $\mathcal{D}$ run over the different types of error generators and combinations of Pauli operators.
As a result, the time-dependent error rate can be extracted from the error generator $L$ via the formula
\begin{equation}
    H_P(t) = \trace[\mathcal{H}_P'^\dag L(t)],~S_P(t) = \trace[\mathcal{S}_P'^\dag L(t)],
\end{equation}
where we represent the superoperators as $d^2 \times d^2$ matrices in the Pauli basis.  

The advantage of changing from the standard Pauli basis to this representation is that many conventional noise sources  can be described as either coherent or stochastic errors.  As a result, we can compress the representation of the error generator from $d^4$ terms to $2(d^2-1)$.  A notable exception is the case of $T_1$ decay processes, but we saw in Sec.~\ref{sec:spinqubits} that these contribute negligibly to the individual gate fidelities.

Similar to the  matrix elements in the Pauli basis, we can compute the power-spectral density of the coefficients of each noise type.  
In Fig.~\ref{fig:errgen}(d), we show an example for a simulated CZ gate using the noise model of Sec.~\ref{sec:exp}.  We plot the power-spectral density of the Hamiltonian and stochastic contributions to the error generators.  We took a different value of $T_2^*$ in Fig.~\ref{fig:errgen}(d) to amplify the effects of the errors for illustrative purposes.  In spin-qubit devices with micromagnets, $T_2^*$ often varies over a wide range depending on the details of the magnetic field gradients and charge noise \cite{Sigillito19}.

As in the single-qubit case, we see that the error generator is dominated by low-frequency fluctuations that dominate over the Markovian noise sources.  This behavior is further borne out by the fidelity benchmarks shown in the inset to Fig.~\ref{fig:errgen}(d).  Moreover, the stochastic noise is strongly suppressed compared to the coherent errors, suggesting that the noise model can be greatly simplified to only include quasi-static coherent noise.  In the following subsections, we show how to use this information to develop effective models for the qubit dynamics to predict the behavior of more complicated RB experiments.

\subsection{Compressed Gate Noise Model}
\label{sec:trunc}

In this sub-section, we use our characterization studies in the previous section to simplify the noise models for quantum circuit dynamics.  Based on the observed power spectral densities for the error generators, we use a quasistatic model for the error generators for each single and two-qubit gate.  As our gate set, we take (1) the identity gate on each site, $\pi/2$-rotations about the (2) $X$ and (3) $Y$ axis, and (4) a two-qubit CZ gate.  We also allow for arbitrary rotations about the $Z$-axis that can be implemented without noise in software \cite{McKay17}.  
We model the error generators for the one and two-qubit gates by Hamiltonian noise
\begin{align*}
    L_{ia} & = \sum_P h_{iaP} \mathcal{H}_P,\\
    L_{CZ} & = h_{XX}(\mathcal{H}_{XX}+\mathcal{H}_{YY})+h_{ZZ}\mathcal{H}_{ZZ} \\ &+h_{ZI}\mathcal{H}_{ZI}+h_{IZ}\mathcal{H}_{IZ},
\end{align*}
where $P$ indexes the Pauli group, $i$ indexes the qubit and $a$ indexes the single-qubit gates.  Each coefficient is modeled as 
\begin{align}
    h_{iaP} &= \bar{h}_{iaP}+\delta h_{iaP} R_a, P\ne I,\\
    h_{PQ} &= \bar{h}_{PQ}+\delta h_{PQ} (R_1+R_2)/\sqrt{2},~P,Q\ne I,\\
    h_{PI/IP} &= \bar{h}_{PI/IP}+\delta h_{PI/IP} R_{1/2},~P\ne I,
\end{align}
 where $R_{1/2}$ are Gaussian random fields drawn randomly at the start of each circuit implementation.  For each set of parameters and gate configurations, we directly measure each coefficient from the full time-dependent simulation of the gate using the noise model described in Sec.~\ref{sec:exp}.  

 As an example, we take the gate parameters from Fig.~\ref{fig:errgen} and, fitting vs $\epsilon = t_g/T_2^*$ for varying values of $T_2^*$, we find a compressed gate model
 \begin{align*}
     \bar{h}_{1P}&=0,~\delta h_{1Z} = 1.4 \epsilon,~\delta h_{1X/Y}=0 \\
     \bar{h}_{2Z}&=\bar{h}_{2Y}=0,~\delta h_{2Z}=\delta h_{2Y}= 0.0034+0.86 \epsilon, \\ \bar{h}_{2X} &= 0.018 - 0.031 \epsilon - 0.18 \epsilon^2, \\
     \delta h_{2X}&=0.018-0.088 \epsilon +0.43 \epsilon^2,\\
     \bar{h}_{3Z}&=\bar{h}_{3X}=0,~\delta h_{3Z}=\delta h_{3X}= 0.0034+0.86 \epsilon, \\ \bar{h}_{3Y} &= 0.018 - 0.031 \epsilon - 0.18 \epsilon^2, \\
     \delta h_{3Y}&=0.018-0.088 \epsilon +0.43 \epsilon^2, \\
     \bar{h}_{XX}&=0.016,~\bar{h}_{ZZ}=-0.006,~\bar{h}_{IZ/ZI}=0,\\
     \delta h_{XX}&=0.0007 + 0.15 \epsilon,h_{ZZ}=0.036,\\
     \delta h_{IZ/ZI}&=0.0009 +1.4 \epsilon .
 \end{align*}
 This model is quantitatively accurate as it includes the dominant contributions to the error generators for each gate, while also accounting for the dominant temporal fluctuations observed in the power-spectral density.   The reason for letting $T_2^*$ vary in this model is that its value often sensitively depends on the details of the magnetic field gradients and charge noise in the system and can vary substantially across spin-qubit devices \cite{Sigillito19}.  In addition, we found that in the noise regime relevant to experiments, $T_2^*$ has the biggest impact on gate performance.

\begin{figure}[t!]
\begin{center}
\includegraphics[width=.45\textwidth]{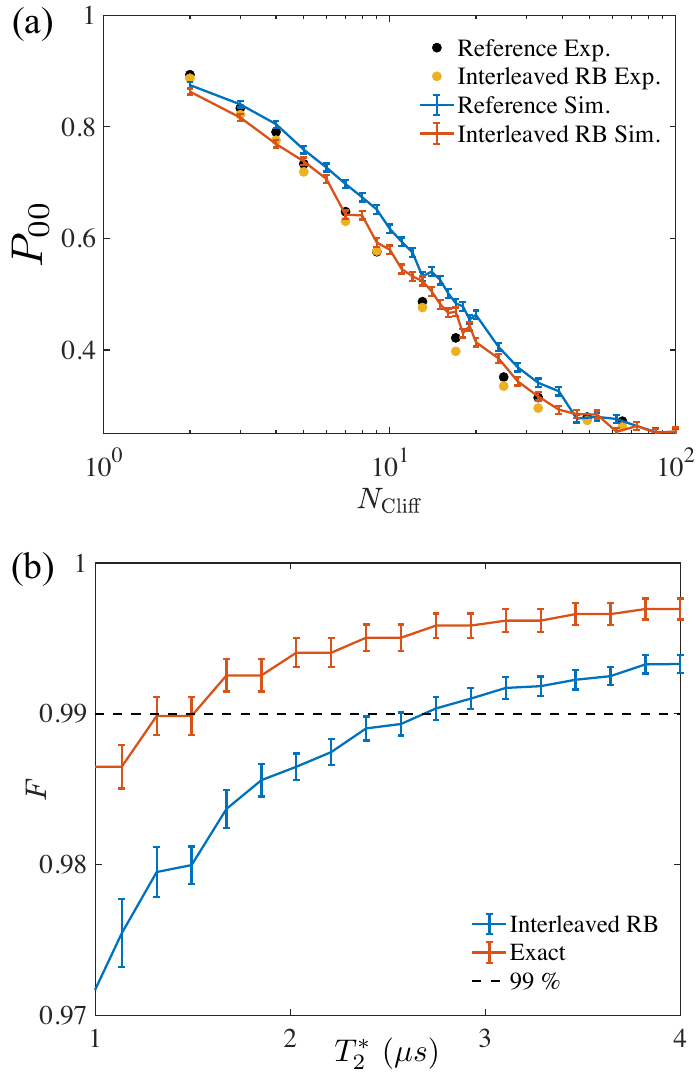}
\caption{(a) Simulated return probability for two-qubit interleaved randomized benchmarking circuits compared to experimental data from Ref.~\cite{Mills22}.  The reference curves correspond to circuits consisting of layers of randomly generated two-qubit Clifford gates, while the interleaved curves include a CZ gate in between each random Clifford operation. We used the parameterized error generators from Sec.~\ref{sec:trunc} with $T_2^* = 3.5 ~\mu$s and included additional two-qubit $ZZ$ errors after each gate to model residual exchange with $h_{\rm ex} = 0.086$.  (b) Comparison between exact average gate fidelity and fidelity obtained from interleaved randomized benchmarking.  The interleaved error rate comes within a factor of two of the exact error rate, but consistently underestimates the true value.}
\label{fig:irb}
\end{center}
\end{figure}

\subsection{Interleaved Randomized Benchmarking}

The compressed gate model introduced in the previous sub-section is particularly amenable to efficient numerical simulations of quantum circuit dynamics.  In contrast to the full time-dependent Hamiltonian simulations, each gate is simulated as a single-time step as opposed to several thousand.  Furthermore, since the errors are purely coherent it is sufficient to evolve pure states.  As a result, it becomes possible to efficiently simulate complex RB circuits that consists of hundreds of individual gates per circuit and many random realizations.  

In Fig.~\ref{fig:irb}(a) we perform a direct simulation of interleaved RB for the two-qubit CZ gate within the compressed gate noise model and compare it to the experimental data from Ref.~\cite{Mills22}.    We generate sequences of random Clifford gates with an inverting gate at the end and plot the probability of returning to the two qubit initial state $\ket{00}$, which leads to the reference decay curves. In the interleaved sequence we insert a CZ gate in between each random Clifford gate \cite{Magesan11}.  From the relative decay rate of each curve we can then extract an estimate of the average two-qubit gate fidelity.  To obtain better quantitative agreement between theory and experiment, we found it important to include a residual exchange interaction term following each single-qubit gate, which we modeled by an additional error channel following the single-qubit gates with the coherent error generator
\begin{equation}
H_{\rm ex} = h_{\rm ex} Z \otimes Z.
\end{equation}
In modeling the data, we took a value $h_{\rm ex} = 0.086$ that translates to a residual exchange on the order of 500~kHz. This value corresponds to the same approximate amount as was measured in the experiment \cite{Mills22}. 

To gain insight into the accuracy of interleaved RB, in Fig.~\ref{fig:irb}(b), we compare the exact CZ gate fidelity against the estimate obtained from interleaved RB for different values of $T_2^*$.  Interleaved RB is simulated using the compressed gate noise model, which includes static gate-dependent coherent noise.  It has been shown that RB leads to systematic errors in the extracted gate fidelities for such gate-dependent noise models \cite{Epstein14,Ball16,Fong17}.  We observe similar behavior in our model, with interleaved RB displaying a systematic deviation from the true fidelity.  Nevertheless, the relative error on the infidelity agrees to within a factor of two and we see that interleaved RB serves to systematically \emph{underestimate} the true fidelity.  As a result, under these conditions we expect the interleaved RB fidelity to serve as a good proxy for the average gate fidelity, lending further theoretical support to recent experimental observations of high-fidelity two-qubit gates in silicon spin qubits based on interleaved RB 
 measurements \cite{Mills22}.

\section{Discussion}
\label{sec:disc}

We developed a systematic framework to address the effects of time-correlated noise in QPT.  Through comparison to experiment, we validated a detailed theoretical model for noise in semiconductor spin qubits.  Extending these results, we showed how to accurately account for such noise processes in QPT.  Using our methods, we were then able to significantly compress the noise model needed to accurately predict noisy quantum circuit evolution.  We then compared predicted interleaved RB fidelities with experimental measurements and the exact process fidelities, finding good overall agreement that supports recent experimental observations of high-fidelity two-qubit operation in silicon spin qubits \cite{Xue22,Noiri22,Mills22}.

Looking to future work, we expect that our methods for QPT will enable a significant speed-up of GST for solid-state qubits using methods based on compressed noise models  \cite{Brieger21}.  The more rapid characterization of qubit operations afforded by these methods will enable more frequent device calibrations and, therefore, the maintenance of high-fidelity operation over longer time scales.  These improvements are crucially needed for fault-tolerant quantum computing.  Moreover, we expect that  systematic studies of fluctuation statistics of the error generators across different platforms, devices and fabrication processes will deepen the collective understanding of relevant noise sources for quantum device operation.  

In the specific context of superconducting qubits, we also expect our approach to have broad applicability.  A key difference in superconducting qubits is the shorter $T_1$ times compared to spin qubits that are often a large source of error in gate operations \cite{Barends19}.  As a result, a compressed noise model for superconducting qubits is less likely to be dominated by coherent errors.  Nevertheless, our analysis of the power-spectral density of the error generators and/or process matrix elements will directly extend over to this case.  In addition, the primary speedup we found in the numerical simulations, arises from switching from a continuous time model to a discrete time model.  Such a speed up would also arise in modeling superconducting qubits.

As a further extension of this research, it is important to investigate more genuinely quantum sources of time-correlated or non-Markovian noise that arise from coherent interactions with an environment.  Nearby quantum-coherent two-level fluctuators and residual nuclear spins could both serve as rich sources of time-correlated noise in spin qubit devices, although nuclear spins typically fluctuate only on timescales much longer than considered in this work \cite{Burkard23}.   Improving characterization methods to account for these quantum-coherent sources of non-Markovian noise is crucial for further validating the operation of solid-state quantum devices.

\begin{acknowledgements}
We thank M. D. Stewart, Jr. and G. White for helpful discussions.
 We acknowledge support from ARO grants W911NF-15-1-0149, W911NF-23-1-0104, and W911NF-23-1-0258.
\end{acknowledgements}

\bibliographystyle{apsrev-nourl-title}
\bibliography{QuantumDot.bib}

\end{document}